\documentclass[10pt]{article}
\usepackage{epsfig}
\usepackage{amsfonts}
\usepackage{amsopn}
\usepackage{amssymb}
\usepackage{amsmath}
\usepackage{graphicx}
\usepackage{amsthm}
\numberwithin{equation}{section}
\usepackage{authblk}
\usepackage[left=1.5cm, right=1.5cm]{geometry}
\newcommand{\NL}{\big(\!\!\big(}
\newcommand{\NR}{\big)\!\!\big)}

\begin{document}

\vspace*{-1.5cm}
\thispagestyle{empty}
\begin{flushright}
AEI-2014-067
\end{flushright}
\vspace*{2.5cm}
\begin{center}
{\Large
{\bf The Lorentz Anomaly via Operator Product Expansion}}
\vspace{2.5cm}

{\large Stefan Fredenhagen\footnote[1]{{\tt E-mail: stefan.fredenhagen@aei.mpg.de}}, Jens Hoppe\footnote[2]{{\tt E-mail: hoppe@kth.se}}, Mariusz Hynek\footnote[3]{{\tt E-mail: mkhynek@kth.se}}}
\vspace*{0.5cm}

$^{1}$Max-Planck-Institut f{\"u}r Gravitationsphysik,
Albert-Einstein-Institut\\
Am M{\"u}hlenberg 1,
14476 Golm, Germany\\
\vspace*{0.5cm}

$^{2,3}$Department of Mathematics, Royal Institute of Technology, KTH\\
100 44 Stockholm, Sweden\\
\vspace*{3cm}

{\bf Abstract}\\[6mm]
\begin{minipage}{14cm}
The emergence of a critical dimension is one of the most striking features of string theory. One way to obtain it is by demanding closure of the Lorentz algebra in the light-cone gauge quantisation, as discovered for bosonic strings more than fourty years ago. We give a detailed derivation of this classical result based on the \emph{operator product expansion} on the Lorentzian world-sheet.

\end{minipage}
\end{center}
\newpage

\tableofcontents
\vspace*{1cm}
\section{Introduction}
More than forty years ago, 26 was noted as a critical dimension
for the dual-resonance models that preceded string theory
\cite{Lovelace:1971fa,Brower:1972wj,Goddard:1972iy}. One
way of obtaining the critical dimension has been to show that in light-cone
gauge quantisation the longitudinal Lorentz-operators $M_{i-}$,
$i=1,\dots ,d>1$, only
commute in $D=d+2=26$ space-time
dimensions~\cite{Goddard:1973qh}. In
this computation the generators $M_{i-}$ are normal-ordered infinite
sums cubic in the oscillator modes of the quantised string.\footnote{cp.\ \cite{GA}; standard textbooks have curiously refrained
from presenting the calculation in detail}
Whereas the quantisation of the string is well understood,
much less is known for general $M$-dimensional extended
objects; however, in~\cite{Hoppe:2010nd} it was noted that, 
as a consequence of Lorentz invariance, a dynamical symmetry exists - which might give a way to algebraically determine the spectrum if one can understand this symmetry in the quantum
theory. Classically, these higher-dimensional objects can be described
-- similarly to the string -- in the light-cone gauge, but the
corresponding world-volume theories are not free as in the case of
strings. Therefore one does not have an expansion in terms of harmonic
oscillators that would make it possible to quantise the theory
directly. On the other hand, one can still use field theory techniques
like operator-product expansions (OPE) in the computations. 

These considerations motivated us to rederive
the critical dimension of bosonic string theory in the light-cone
gauge quantisation by only using the operator product expansion on the
Lorentzian world-sheet. As the world-sheet theory is free, the OPEs
are simple and completely equivalent to the harmonic oscillator
commutators of the corresponding modes (so that it is guaranteed that the
result is the same as in the oscillator approach). The computation, however, turned out
to be surprisingly tedious and subtle (it involves a careful treatment of
composite and non-local operators). We decided to write it
up and present it in this note, in the hope that the approach might be useful for
higher dimensional extended objects (note also~\cite{Marquard:1988bj}), - as well as an alternative derivation of the critical dimension (see
also~\cite{Bering:2011uf} where yet another derivation of the Lorentz
anomaly was presented). We should add that, similar to the oscillator computation, 
it seems that one can not pinpoint any particular step of the computation where the anomaly
arises; it rather appears as a result of an interplay
of several anomalous terms that arise due to the regularisations
needed to define the composite operators. Note however 
that, because we work on the (Lorentzian) cylinder, there is no need
to artificially introduce a normal-ordering constant in the
computation; it is already set to the right value by using the
most natural definition of composite operators.

Let us describe the computation in a language that can also be used for
higher-dimensional extended objects (see ~\cite{Hoppe:2011dj} for some
naive heuristic considerations). The string is parameterised by a
map from the Lorentzian cylinder (time coordinate $t$ and angular
variable $\varphi$) to a flat Minkowski space. In the light-cone gauge
the degrees of freedom are carried by the transversal fields $\vec{x}$ and their
conjugate momenta $\vec{p}$, as well as by the zero mode $\zeta_{0}$
of the coordinate $\zeta=x_{-}$ and by its conjugate variable $\eta$.

Classically the longitudinal generators of the Lorentz algebra are
given by
\begin{equation}
M_{i-}:=\int_0^{2 \pi}(x_i\mathcal{H}- \zeta p_i)d \varphi,~~
i=1,\dotsc ,d=D-2 \ .
\end{equation}
Here
\begin{equation}
\mathcal{H}( \varphi) :=
\frac{\pi}{\eta}(\vec{p}\cdot\vec{p}+\vec{x}'\cdot \vec{x}')
\label{classham}
\end{equation} 
is the classical Hamiltonian density (corresponding to $p_{-}$), and 
\begin{equation}
\zeta(\varphi)=\zeta_0-\frac{2 \pi}{\eta} \int_{\varphi}^{2 \pi}
\vec{p}\cdot\vec{x}'\,d \psi+\frac{1}{\eta} \int_{0}^{2 \pi}
\vec{p}\cdot\vec{x}'\,\psi d\psi+\frac{\pi - \varphi}{\eta} \int_{0}^{2
\pi} \vec{p}\cdot\vec{x}'\,d \psi
\label{zeta}
\end{equation}
is the reconstructed $x_{-}$ coordinate of the string that follows
from $\zeta'= \frac{2 \pi}{\eta} \vec{p}\cdot\vec{x}'$. The transversal
fields $x_i(\varphi), p_j(\varphi)$ are constrained by
\begin{equation}
\int_0^{2 \pi} \vec{p}\cdot\vec{x}' d \varphi=0 \ ,
\label{constr}
\end{equation}
so that the last term in (\ref{zeta}) could be dropped, and $\zeta$
can be rewritten as 
\begin{equation}
\zeta(\varphi)=\zeta_0+\frac{2 \pi}{\eta} \int_{0}^{\varphi}
\vec{p}\cdot\vec{x}'\,d \psi+\frac{1}{\eta} \int_{0}^{2 \pi}
\vec{p}\cdot\vec{x}'\,\psi d\psi \ .
\label{zeta3}
\end{equation}
That the $M_{i-}$ Poisson-commute  (provided (\ref{constr}) holds)
is a particular case of a result of Goldstone~\cite{Goldstone:1985}, who for
arbitrary dimension $M$ of the extended object solved
\begin{equation}
\frac{\partial \zeta}{\partial \varphi^a}= \frac{\vec{p}\cdot\partial_a
\vec{x}}{\eta \rho}, ~~a=1,\dotsc ,M
\label{goldeq}
\end{equation}
for $\zeta$ in terms of $\vec{x}$ and $\vec{p}$ and some Green's
function $G$ ($\rho$ is a density satisfying $\int \rho \,d^M
\varphi=1$) and then showed that classically, for all $M$, the generators of the inhomogeneous
Lorentz group can be consistently realised on the $(\eta, \zeta_0,
\vec{x}(\varphi), \vec{p}(\varphi))$ phase-space constrained by the
consistency of~(\ref{goldeq}) (which for $M=1$ and $\rho=\frac{1}{2
\pi}$ simply becomes~(\ref{constr})). In the string case considered here,
\begin{equation}
G(\varphi,\psi)=2 \pi (\psi -
\varphi)\theta(\psi-\varphi)-\frac{1}{2}(\psi-\varphi+\pi)^2+\frac{\pi^2}{6}\ ,
\end{equation}
and the reconstructed $x_{-}$ coordinate is
\begin{equation}
\zeta(\varphi)=\zeta_0-\frac{1}{\eta}\int_0^{2 \pi}
\partial_{\psi}G(\varphi,\psi)\,\vec{p}(\psi)\cdot \vec{x}'(\psi)\,d
\psi \ ,
\end{equation}
which reduces to the expression~\eqref{zeta3} given above.

The paper is organised as follows. In section~\ref{sec:OPE} we explain
the OPE techniques that are needed to get to a quantum definition of
the Lorentz operators. We then derive the commutation relations of all
basic fields in section~\ref{sec:commutationrelations}. Finally, we
compute the crucial commutator $[M_{i-},M_{j-}]$ in
section~\ref{sec:crucial}. The three appendices contain some
technical parts of the computation.

\section{Operator product expansion and composite operators}
\label{sec:OPE}
The definition of the Lorentz generators involves products of fields
which we have to define properly in the quantum theory. In a free
theory this can be done by using an oscillator expansion of the free
fields and then define normal-ordered products by moving annihilation
operators to the right of creation operators. Alternatively we can use
the operator product expansion (OPE) of the fields to define composite operators
by subtracting the singular part of the OPE. This leads to
an equivalent description for free fields, but it can in principle also be used in
more general situations where the usual normal ordering prescription
in terms of annihilation and creation operators is not possible.

In the case at hand, the $x_{i}$ are massless free fields on the
two-dimensional cylinder, and their OPE reads (no summation
over $i$)
\begin{equation}
x_{i} (\tilde{\varphi}) x_{i} (\varphi) = - \frac{1}{2\pi} \log \big| \sin
\tfrac{\tilde{\varphi}-\varphi}{2}\big| + \text{regular}\ .
\end{equation}
Then the OPE of the fields $x_{i}'$ is given by
\begin{equation}\label{xprimeOPE}
x'_{i} (\tilde{\varphi}) x'_{i} (\varphi)
= S_{\text{sing}} (\tilde{\varphi},\varphi)
+ \text{regular}\ ,
\end{equation}
with the distribution $S_{\text{sing}}$ in two variables $\tilde{\varphi}$ and
$\varphi$ given by
\begin{equation}
S_{\text{sing}} (\tilde{\varphi},\varphi) =
-\frac{1}{2\pi}\partial_{\tilde{\varphi}}\partial_{\varphi} \log \big| \sin
\tfrac{\tilde{\varphi}-\varphi}{2}\big| = 
\frac{1}{4 \pi}
\partial_{\tilde{\varphi}}\left(\cos\Big(\frac{\tilde{\varphi}-\varphi}{2}\Big)\mathcal{P}\frac{1}{\sin(\frac{\tilde{\varphi}-\varphi}{2})}\right)
\ ,
\end{equation}
where $\mathcal{P}$ denotes the principal value. To define the operator ''$x_{i}'(\varphi)x_{i}'(\varphi)$'' we use
point-splitting, so we evaluate~\eqref{xprimeOPE} for
$\tilde{\varphi}=\varphi-\epsilon$ and determine the singular piece,
\begin{equation}
x'_{i} (\tilde{\varphi}) x'_{i}
(\varphi)\Big|_{\tilde{\varphi}=\varphi-\epsilon} 
= -\frac{1}{2\pi \epsilon^{2}} + \text{regular} \ .
\end{equation}
Note that away from $\varphi=\tilde{\varphi}$, $S_{\text{sing}}$ is a regular
function, and we can replace $\tilde{\varphi}=\varphi-\epsilon$. This
singular piece is then subtracted to define the product of $x_{i}'$
with itself,
\begin{equation}
\NL x'_{i}x'_{i} \NR  (\varphi) := \lim_{\epsilon\to 0} \left( x'_{i}
(\tilde{\varphi}) x'_{i}
(\varphi)\Big|_{\tilde{\varphi}=\varphi-\epsilon}
+\frac{1}{2\pi\epsilon^{2}}\right)\ .
\end{equation}
Up to an additive constant this is equivalent to the normal ordering
prescription using oscillators.

Similarly we have
\begin{equation}\label{ppOPE}
p_i(\tilde{\varphi})p_i(\varphi)= S_{\text{sing}} (\tilde{\varphi},\varphi)
+ \text{regular}\ ,
\end{equation}
and
\begin{equation}
\NL p_{i}p_{i} \NR (\varphi) := \lim_{\epsilon\to 0} 
\left(
p_i(\tilde{\varphi})p_i(\varphi)\Big|_{\tilde{\varphi}=\varphi-\epsilon}
+ \frac{1}{2\pi \epsilon^{2}}\right) \ .
\end{equation}
This then leads to the quantum definition of $\mathcal{H}$,
\begin{align}
\mathcal{H}(\varphi) &= \frac{\pi}{\eta}\sum_{i}\left(\NL p_{i}p_{i}
\NR  (\varphi)
+ \NL x'_{i}x'_{i} \NR  (\varphi) \right)\\
&= \lim_{ \epsilon \rightarrow 0} \frac{\pi}{\eta}\left(\sum_{i}\left(
p_i(\tilde{\varphi})p_i(\varphi)+x'_i(\tilde{\varphi})x'_i(\varphi)\right)\Big|_{\tilde{\varphi}=\varphi-\epsilon}
+ 2(D-2) \frac{1}{2\pi\epsilon^{2}} \right) \ .
\label{QMdefH}
\end{align}
In the definition of $\zeta$ we also meet the product of $p_{i}$ and
$x_{i}'$. Their operator product expansion only has singularities of
contact type,
\begin{equation}
p_i(\tilde{\varphi})x'_i(\varphi)  = \frac{i}{2}
\partial_{\tilde{\varphi}}\delta(\tilde{\varphi}-\varphi) + \text{regular} \ .
\end{equation}
We therefore get a well-defined composite operator just by
point-splitting,
\begin{equation}
\NL p_{i}x_{i}'\NR (\varphi) = \lim_{\epsilon\to 0}
\left(p_{i}(\tilde{\varphi}) x_{i}' (\varphi)
\right)\Big|_{\tilde{\varphi}=\varphi-\epsilon} \ ,
\end{equation}
and we can define the quantum version of $\zeta$ as
\begin{equation}
\zeta (\varphi) = \zeta_0-\frac{1}{\eta}\int_0^{2 \pi}
\partial_{\psi}G(\varphi,\psi)\NL\vec{p}\cdot \vec{x}'\NR (\psi)\,d
\psi \ .
\end{equation}
When we define the Lorentz generators $M_{i-}$ we also encounter the
product of $\mathcal{H}$ and $x_{i}$ as well as the product of $\zeta$
and $p_{i}$, which we have to regularise to obtain well-defined expressions.

Let us start with the product of $\mathcal{H}$ and $x_{i}$. The
singularities in the operator product expansion follow via Wick's
theorem from the individual contractions of $x_{i}$ and the $x_{j}'$
appearing inside $\mathcal{H}$,
\begin{equation}\label{xHOPE}
x_{i}(\tilde{\varphi}) \mathcal{H}(\varphi)\Big|_{\tilde{\varphi}=\varphi-\epsilon}
= \frac{\pi}{\eta} \left(-\frac{1}{\pi\epsilon} x_{i}'(\varphi)\right)
+ \text{regular} \ .
\end{equation}
Therefore we can define the quantum product of $x_{i}$ and
$\mathcal{H}$ by
\begin{equation}
\NL x_{i} \mathcal{H} \NR (\varphi) = \lim_{\epsilon\to 0} \left(
x_{i}(\tilde{\varphi}) \mathcal{H}(\varphi)\Big|_{\tilde{\varphi}=\varphi-\epsilon}
+ \frac{1}{\epsilon\eta} x_{i}'(\varphi) \right) \ .
\end{equation}
A little more work is needed to define the product of $p_{i}$ and
$\zeta$, because $\zeta$ is defined as a non-local expression in the
fields. The possible singularities come from the contact
singularity between $p_{i}$ and $x_{j}'$ inside $\zeta$, and from the
singularity between $p_{i}$ and the $p_{j}$ inside $\zeta$. The contact singularity
is avoided if we consider the symmetrised product $p_{i}\zeta + \zeta
p_{i}$, and we find
\begin{equation}
\frac{1}{2}\left(p_{i}(\tilde{\varphi}) \NL \vec{p}\cdot \vec{x}'\NR (\psi) + 
 \NL \vec{p}\cdot \vec{x}'\NR (\psi) \,p_{i}(\tilde{\varphi}) \right)
= -\frac{1}{4\pi}x_{i}'(\psi) \,\partial_{\psi}\left(\cos\Big(\frac{\psi-\tilde{\varphi}}{2}\Big)\mathcal{P}\frac{1}{\sin(\frac{\psi-\tilde{\varphi}}{2})}\right)
+ \text{regular} \ .
\end{equation}
The possible singularity in the symmetrised product of $\zeta$ and
$p_{i}$ is then 
\begin{align}
&\frac{1}{2}\left(
p_{i}(\tilde{\varphi})\zeta(\varphi)+\zeta(\varphi)p_{i}(\tilde{\varphi})\right)\Big|_{\tilde{\varphi}=\varphi-\epsilon}\nonumber\\
&\qquad \qquad = \frac{1}{4\pi\eta} \int_{0}^{2\pi} \partial_{\psi}G(\varphi,\psi)
\,x_{i}'(\psi)\,
\partial_{\psi}\left(\cos\Big(\frac{\psi-\tilde{\varphi}}{2}\Big)\mathcal{P}\frac{1}{\sin(\frac{\psi-\tilde{\varphi}}{2})}\right)\Big|_{\tilde{\varphi}=\varphi-\epsilon}d\psi 
+ \text{regular} \\
&\qquad \qquad = -\frac{1}{4\pi\eta} \int_{0}^{2\pi} x_{i}(\psi)\, \partial_{\psi}\left(\partial_{\psi}G(\varphi,\psi)\,
\partial_{\psi}\left(\cos\Big(\frac{\psi-\tilde{\varphi}}{2}\Big)\mathcal{P}\frac{1}{\sin(\frac{\psi-\tilde{\varphi}}{2})}\right)\Big|_{\tilde{\varphi}=\varphi-\epsilon}\right) d\psi 
+ \text{regular} \\
&\qquad \qquad =  \partial_{\varphi}\left(\frac{1}{4\pi\eta} \int_{0}^{2\pi} x_{i}(\psi) \left(\partial_{\psi}G(\varphi,\psi)\,
\partial_{\psi}\left(\cos\Big(\frac{\psi-\tilde{\varphi}}{2}\Big)\mathcal{P}\frac{1}{\sin(\frac{\psi-\tilde{\varphi}}{2})}\right)\Big|_{\tilde{\varphi}=\varphi-\epsilon}\right)\right) d\psi 
+ \text{regular}
\ .
\end{align}
The possible singular part is therefore a total derivative in
$\varphi$, which means that it does not matter in the expression for
$M_{i-}$, which involves an integration over $\varphi$.

Similarly, also the singular part of the product of $x_{i}$ and
$\mathcal{H}$ is a total derivative (see~\eqref{xHOPE}), which
vanishes upon integration. Therefore the quantum definition of
$M_{i-}$ using symmetrised products and point-splitting is given by
\begin{align}
M_{i-} &= \lim_{\epsilon, \delta \rightarrow 0}
M_{i-}(\epsilon, \delta) \\
&= \lim_{\epsilon, \delta \rightarrow 0}
\frac{1}{2} \int \big(x_i(\varphi+\epsilon) \mathcal{H}(\varphi)+ \mathcal{H}(\varphi)x_i(\varphi+\epsilon)-\zeta(\varphi+\delta)p_i(\varphi)-
p_i(\varphi)
\zeta(\varphi+\delta)\big) \, d\varphi \ .
\label{defofM}
\end{align}
This is our starting point for analysing the commutator of $M_{i-}$ and
$M_{j-}$.

\section{Basic commutation relations}
\label{sec:commutationrelations}
To compute the commutators of the Lorentz algebra generators we
need to determine the commutators of the fields
$x_i(\varphi),p_j(\varphi), \mathcal{H}(\varphi)$, $\zeta(\varphi)$,
which follow from the canonical commutation relations of $x_{i}$ and
$p_{j}$, and $\eta$ and $\zeta_{0}$.
We first list the results, and present the
derivation subsequently. The commutators are
\begin{align}
[\eta ,\zeta_{0}] &= i \label{etazeta}\\
[x_i(\varphi),p_j(\tilde{\varphi})]&=i \delta_{ij}\,\delta(\varphi-\tilde{\varphi})\label{xp}\\
[\mathcal{H}(\varphi),p_j(\tilde{\varphi})]&=
\frac{2 \pi i}{\eta}\partial_{\varphi}\delta(\varphi-\tilde{\varphi})\,x_j'(\varphi)\label{Hp}\\
[x_i(\varphi),\zeta(\tilde{\varphi})]&=-\frac{i}{\eta}\partial_{\varphi}G(\tilde{\varphi},\varphi)\,x_i'(\varphi)\label{xzeta}\\
[\mathcal{H}(\varphi),\zeta(\tilde{\varphi})]&=-\frac{2 \pi i}{\eta}\delta(\varphi-\tilde{\varphi})\,\mathcal{H}(\varphi)-\frac{i}{\eta}\partial_{\varphi}\Big(\partial_{\varphi}G(\tilde{\varphi},\varphi)\mathcal{H}(\varphi)\Big)+\frac{\pi i}{3 \eta^2}(D-2)\,\partial_{\varphi}^2\delta(\varphi-\tilde{\varphi})\label{Hzeta}\\
[\mathcal{H}(\varphi),\mathcal{H}(\tilde{\varphi})]&=\frac{2 i
\pi^2}{\eta^2}\partial_{\varphi}\delta(\varphi-\tilde{\varphi})\Big(\vec{p}(\varphi)\cdot\vec{x}'(\tilde{\varphi})+
\vec{x}'(\tilde{\varphi})\cdot\vec{p}(\varphi)+\vec{p}(\tilde{\varphi})\cdot\vec{x}'(\varphi)+\vec{x}'(\varphi)\cdot\vec{p}(\tilde{\varphi})\Big)\label{HH}\\
[\mathcal{H}(\varphi),x_j(\tilde{\varphi})]&=-\frac{2 \pi i}{\eta}\delta(\varphi-\tilde{\varphi})\,p_j(\tilde{\varphi})\label{Hx}\\
[\zeta(\varphi),p_j(\tilde{\varphi})]&=\frac{i}{\eta}\partial_{\tilde{\varphi}}\Big(\partial_{\tilde{\varphi}}G(\varphi,\tilde{\varphi})\,p_j(\tilde{\varphi})\Big)\label{zetapkom}\\
[\zeta(\varphi),\zeta(\tilde{\varphi})]&=-\frac{i}{\eta}\partial_{\varphi}G(\tilde{\varphi},\varphi)
\Big(\zeta'(\varphi) + \zeta'(\tilde{\varphi})\Big) \ .
\label{zeta2}
\end{align}
The derivation is straightforward:
\begin{itemize}
\item (\ref{etazeta}),(\ref{xp}): These are the canonical commutation relations.
\item (\ref{Hp}): We use the definition of $\mathcal{H}$ via
point-splitting (see~\eqref{QMdefH}). The only contribution to the commutator
comes from the $x'^{2}$ term in $\mathcal{H}$,
\begin{align}
[\mathcal{H}(\varphi),p_j(\tilde{\varphi})]&= \frac{\pi}{\eta}\lim_{\epsilon \rightarrow 0} \big[\vec{x}'(\varphi)\cdot \vec{x}'(\varphi-\epsilon),p_j(\tilde{\varphi}) \big]\\
&=
 i \frac{\pi}{\eta} \lim_{\epsilon \rightarrow 0}\Big(x_j'(\varphi)\,
 \partial_{\varphi} \delta(\varphi - \tilde{\varphi} -
 \epsilon)+x_j'(\varphi+\epsilon) \,\partial_{\varphi} \delta(\varphi -
 \tilde{\varphi})\Big)= \frac{2 \pi i}{\eta} \partial_{\varphi} \delta(\varphi - \tilde{\varphi}) \,x_j'(\varphi)\ .
\end{align}
\item (\ref{xzeta}): 
\begin{equation}
[x_i(\varphi),\zeta(\tilde{\varphi})]=\lim_{\epsilon \rightarrow
0}\frac{1}{\eta}\int G(\tilde{\varphi},\psi)\,\partial_{\psi}
\big[x_i(\varphi), \vec{p}(\psi) \cdot \vec{x}'(\psi-\epsilon)\big]d
\psi=-\frac{i}{\eta}\partial_{\varphi}G(\tilde{\varphi},\varphi)\,x_i'(\varphi)
\ ,
\end{equation}
where in the second step we integrated by parts, computed the commutator
$\big[x_i(\varphi), \vec{p}(\psi) \cdot \vec{x}'(\psi-\epsilon)\big]$ and took the limit $\epsilon \rightarrow 0$.
\item (\ref{Hzeta}): The derivation of the commutator of $\mathcal{H}$ and $\zeta$ is
straightforward, but slightly more involved,
\begin{align}
[\mathcal{H}(\varphi),\zeta(\tilde{\varphi})]&=
\bigg[\frac{\pi}{\eta} \Big( \NL \vec{p} \cdot\vec{p}\NR (\varphi)+ \NL \vec{x}' \cdot\vec{x}'\NR (\varphi)\Big), \zeta_0-\frac{1}{\eta} \int \partial_{\psi}G(\tilde{\varphi}, \psi)\NL \vec{p}\cdot \vec{x}'\NR (\psi) \,d\psi\bigg]\\
&=-\frac{i\pi}{\eta^2} \Big(\NL \vec{p} \cdot\vec{p}\NR (\varphi)+\NL \vec{x}' \cdot\vec{x}'\NR (\varphi)\Big)\nonumber\\
&\qquad -\frac{\pi}{\eta^{2}} \lim_{\epsilon, \tilde{\epsilon} \rightarrow 0} \int \partial_{\psi}G(\tilde{\varphi}, \psi)\big[\vec{p}(\varphi) \cdot\vec{p}(\varphi-\epsilon)+\vec{x}'(\varphi) \cdot\vec{x}'(\varphi-\epsilon),\vec{p}(\psi)\cdot \vec{x}'(\psi-\tilde{\epsilon})\big] d\psi\\
&=-\frac{i}{\eta}\mathcal{H}(\varphi) +\frac{2\pi i}{\eta^{2}}
\lim_{\tilde{\epsilon}\rightarrow 0} \int
\partial_{\psi}G(\tilde{\varphi}, \psi)
\Big(\partial_{\psi}\delta (\varphi-\psi+\tilde{\epsilon}) \,\vec{p}(\varphi)\cdot\vec{p}(\psi)
+\partial_{\psi}\delta(\varphi-\psi)\,
\vec{x}'(\varphi)\cdot\vec{x}'(\psi-\tilde{\epsilon})\Big)d\psi \\
&=-\frac{i}{\eta}\mathcal{H}(\varphi) +\frac{2\pi i}{\eta^{2}}
\int \partial_{\psi}G(\tilde{\varphi}, \psi)\, 
\partial_{\psi}\delta (\varphi-\psi) \Big(
\vec{p}(\varphi)\cdot\vec{p}(\psi) + 
\vec{x}'(\varphi)\cdot\vec{x}'(\psi) -
2(D-2)\,S_{\text{sing}}(\varphi,\psi ) \Big)d\psi \nonumber\\
&\qquad +\frac{2 \pi i}{\eta^2}(D-2)\lim_{\tilde{\epsilon} \rightarrow 0} \int \partial_{\psi} G(\tilde{\varphi},\psi)\Big(\partial_{\psi}\delta(\varphi - \psi)\,S_{\text{sing}}(\varphi,\psi-\tilde{\epsilon})+\partial_{\psi}\delta(\varphi-\psi+\tilde{\epsilon})\,S_{\text{sing}}(\varphi,\psi)\Big)d \psi\\
&=-\frac{i}{\eta}\mathcal{H}(\varphi) -\frac{2\pi i}{\eta^{2}}\bigg(\partial_{\varphi}^{2}G (\tilde{\varphi},\varphi) \bigg(\frac{\eta}{\pi}\mathcal{H}(\varphi)+\frac{D-2}{12\pi} \bigg) + \partial_{\varphi}G (\tilde{\varphi},\varphi)\frac{\eta}{2\pi}\partial_{\varphi}\mathcal{H} (\varphi)\bigg)\nonumber\\
&\qquad -\frac{2\pi i}{\eta^{2}} (D-2) \int \partial_{\psi}G (\tilde{\varphi},\psi) \frac{1}{12\pi}\Big(\partial_{\psi}\delta (\varphi-\psi) + \partial_{\psi}^{3}\delta (\varphi-\psi) \Big)d\psi \\
&=-\frac{2 \pi
i}{\eta}\delta(\varphi-\tilde{\varphi})\,\mathcal{H}(\varphi)-\frac{i}{\eta}\partial_{\varphi}\Big(\partial_{\varphi}G(\tilde{\varphi},\varphi)\,\mathcal{H}(\varphi)\Big)+\frac{\pi
i}{3 \eta^2}(D-2)\,\partial_{\varphi}^2\delta(\varphi-\tilde{\varphi}) \ ,
\end{align}
where we used that
\begin{align}
\partial_{\psi}\delta(\varphi - \psi)S_{\text{sing}}(\varphi,\psi-\tilde{\epsilon})&=\partial_{\psi}\delta(\varphi - \psi)S_{\text{sing}}(\varphi,\varphi-\tilde{\epsilon})-\delta(\varphi - \psi)\partial_{\psi}S_{\text{sing}}(\varphi,\psi-\tilde{\epsilon})|_{\psi=\varphi}\\
&=\Big( -\frac{1}{2 \pi \tilde{\epsilon}^2}-\frac{1}{24\pi}+
\mathcal{O}\big(\tilde{\epsilon}^2\big)\Big)\partial_{\psi}\delta(\varphi - \psi)+\Big(\frac{1}{\pi
\tilde{\epsilon}^3}+ \mathcal{O}(\tilde{\epsilon})\Big)\delta(\varphi - \psi) \ .
\end{align}
Note that the commutator of $\mathcal{H}$ and $\zeta$ contains a term
that depends on the number $D$ of space-time dimensions.
\item (\ref{HH}):
\begin{align}
[\mathcal{H}(\varphi),\mathcal{H}(\tilde{\varphi})] &= \frac{\pi^2}{\eta^2}\lim_{\epsilon, \tilde{\epsilon} \rightarrow 0}\big[\vec{p}(\varphi) \cdot \vec{p}(\varphi - \epsilon)+\vec{x}'(\varphi) \cdot \vec{x}'(\varphi - \epsilon), \vec{p}(\tilde{\varphi}) \cdot \vec{p}(\tilde{\varphi} - \tilde{\epsilon})+\vec{x}'(\tilde{\varphi}) \cdot \vec{x}'(\tilde{\varphi} - \tilde{\epsilon})\big]\\
&=\frac{2 i
\pi^2}{\eta^2}\partial_{\varphi}\delta(\varphi-\tilde{\varphi})\Big(\vec{p}(\varphi)\cdot \vec{x}'(\tilde{\varphi})+
\vec{x}'(\tilde{\varphi})\cdot
\vec{p}(\varphi)+\vec{p}(\tilde{\varphi})\cdot
\vec{x}'(\varphi)+\vec{x}'(\varphi) \cdot  \vec{p}(\tilde{\varphi})\Big)\ .
\end{align}
\item (\ref{Hx}): 
\begin{align}
[\mathcal{H}(\varphi),x_j(\tilde{\varphi})]&= \frac{\pi}{\eta}\lim_{\epsilon \rightarrow 0} \big[\vec{p}(\varphi)\cdot \vec{p}(\varphi-\epsilon),x_j(\tilde{\varphi}) \big]\\
&=-i \frac{\pi}{\eta} \lim_{\epsilon \rightarrow 0}\Big(p_j(\varphi)\,
\delta(\varphi - \tilde{\varphi} - \epsilon)+p_j(\varphi+\epsilon)\,
\delta(\varphi - \tilde{\varphi})\Big)= -\frac{2 \pi i}{\eta}  \delta(\varphi - \tilde{\varphi}) \,p_j(\varphi)\ .
\end{align}
\item (\ref{zetapkom}): 
\begin{equation}
[\zeta(\varphi),p_j(\tilde{\varphi})]=-\lim_{\epsilon \rightarrow
0}\frac{1}{\eta}\int G(\varphi,\psi)\,\partial_{\psi}
\big[p_j(\tilde{\varphi}), \vec{p}(\psi) \cdot
\vec{x}'(\psi-\epsilon)\big]d
\psi=\frac{i}{\eta}\partial_{\tilde{\varphi}}\Big(\partial_{\tilde{\varphi}}G(\varphi,\tilde{\varphi})\,p_j(\tilde{\varphi})\Big)
\end{equation}
where in the second equality we integrated by parts, computed the commutator $[p_j(\tilde{\varphi}), \vec{p}(\psi) \cdot \vec{x}'(\psi-\epsilon)]$, integrated by parts again and finally took the limit $\epsilon \rightarrow 0$.
\item (\ref{zeta2}):
\begin{align}
[\zeta(\varphi),\zeta(\tilde{\varphi})]&=\frac{i}{\eta}(\zeta(\varphi)-\zeta(\tilde{\varphi}))\nonumber\\
&\qquad +\frac{1}{4\eta^2}\lim_{\epsilon, \tilde{\epsilon} \rightarrow 0}\int \int \partial_{\tilde{\psi}}G(\tilde{\varphi},\tilde{\psi})\,\partial_{\psi}G(\varphi,\psi)\big[\vec{p}(\psi) \cdot \vec{x}'(\psi-\epsilon), \vec{p}(\tilde{\psi}) \cdot \vec{x}'(\tilde{\psi}-\tilde{\epsilon})\big] d\psi \,d\tilde{\psi}\\
&=\lim_{\epsilon, \tilde{\epsilon} \rightarrow 0} \left(-\frac{2\pi}{\eta^2} \partial_{\varphi}G(\tilde{\varphi},\varphi)\,\vec{p}(\varphi) \cdot \vec{x}'(\varphi-\epsilon-\tilde{\epsilon})-(\varphi \leftrightarrow \tilde{\varphi})\right)\\
&=-\frac{i}{\eta}\partial_{\varphi}G(\tilde{\varphi},\varphi)\big(
\zeta'(\varphi) + \zeta'(\tilde{\varphi})\big) \ ,
\end{align}
where in the last step we used that $\zeta'=\frac{2\pi}{\eta}\NL \vec{p}\cdot \vec{x}'\NR$. 
\end{itemize}

\section{The crucial commutator}
\label{sec:crucial}
We now want to analyse the commutator of the generators $M_{i-}$ and
$M_{j-}$, which we defined in~\eqref{defofM}. It can be decomposed as
\begin{align}
[M_{i-},M_{j-}] &= \lim_{\epsilon, \delta \rightarrow 0}\lim_{\tilde{\epsilon}, \tilde{\delta} \rightarrow 0} \big[M_{i-}(\epsilon, \delta),M_{i-}(\tilde{\epsilon}, \tilde{\delta})\big]\\
&= \lim_{\epsilon, \delta \rightarrow 0}\lim_{\tilde{\epsilon},
\tilde{\delta} \rightarrow 0}\int \int \Big[
x_i(\varphi+\epsilon) \mathcal{H}(\varphi)- \tfrac{1}{2}\big(\zeta(\varphi+\delta)p_i(\varphi)+
p_{i} (\varphi)\zeta(\varphi+\delta)\big),\nonumber\\
&\qquad \qquad \qquad \qquad \quad  x_j(\tilde{\varphi}+\tilde{\epsilon})
\mathcal{H}(\tilde{\varphi})-\tfrac{1}{2}\big(\zeta(\tilde{\varphi}+\tilde{\delta})p_j(\tilde{\varphi})
+p_{j} (\tilde{\varphi})\zeta(\tilde{\varphi}+\tilde{\delta})\big)\Big] d\varphi\,d\tilde{\varphi}\ .
\end{align}
Here we used the fact that $x_i(\varphi+\epsilon)$ and
$\mathcal{H}(\varphi)$ commute for $\epsilon >0$
(see~\eqref{Hx}). There are four types of contributions: the
commutator of the terms of the form $x\mathcal{H}$, the two mixed
commutators of $x\mathcal{H}$ and $\zeta p$, and the commutator of the
terms of the form $\zeta p$.

At the end we want to analyse the behaviour when $\epsilon,\delta$ and
$\tilde{\epsilon},\tilde{\delta}$ go to zero. It is not guaranteed that this limit
exists, and indeed we will see that e.g.\ the commutator of the terms of the
form $x\mathcal{H}$ alone is singular when the regularisation
parameters go to zero; this singularity will go away when we combine
all contributions to the commutator of $M_{i-}$ and $M_{j-}$. On the
other hand we expect that we can take one set of parameters to zero without getting
a singularity in the different contributions: because the individual entries entering the commutator are regularised
and do not show any singularity in $\epsilon,\delta$ or in
$\tilde{\epsilon},\tilde{\delta}$, the only way a new singularity can appear is through terms that become
singular when both type of parameters go to zero (like $(\epsilon+\tilde{\epsilon})^{-1}$). 

Our strategy will therefore be to always take the limit
$\tilde{\epsilon},\tilde{\delta}$ to zero first, and then consider the limit
when $\epsilon$ and $\delta$ go to zero.

We start by analysing the commutator of
the terms $x\mathcal{H}$ in detail. We then present the results for the
remaining commutators, and evaluate the total expression.

\subsection{Commutators of the form $[x_{i}\mathcal{H},x_{j}\mathcal{H}]$}

We want to analyse
\begin{equation}
C_{ij}^{x\mathcal{H},x\mathcal{H}}(\epsilon,\tilde{\epsilon}) = \int \int
\Big[
x_i(\varphi+\epsilon) \mathcal{H}(\varphi) , x_j(\tilde{\varphi}+\tilde{\epsilon})
\mathcal{H}(\tilde{\varphi})\Big] d\varphi \, d\tilde{\varphi}
\end{equation}
by using the commutation relations that we worked out in
section~\ref{sec:commutationrelations}. In a first step we obtain
\begin{align}
C_{ij}^{x\mathcal{H},x\mathcal{H}}(\epsilon,\tilde{\epsilon}) & = 
\frac{2\pi i}{\eta} \int \Big(x_j(\varphi+\tilde{\epsilon}) p_i(\varphi)\mathcal{H}(\varphi-\epsilon)
-x_i(\varphi+\epsilon+\tilde{\epsilon})
p_j(\varphi+\tilde{\epsilon})\mathcal{H}(\varphi) \nonumber\\
&\qquad \qquad \quad +
\big(x_i(\varphi+\epsilon)x'_j(\varphi+\tilde{\epsilon})-x_i'(\varphi+\epsilon)x_j(\varphi+\tilde{\epsilon})\big)\,
\zeta'(\varphi)\Big) d \varphi \ .
\end{align}
We now consider the behaviour when $\tilde{\epsilon}$ goes to zero,
and we find
\begin{align}
C_{ij}^{x\mathcal{H},x\mathcal{H}}(\epsilon,\tilde{\epsilon}) & = 
\frac{2\pi i}{\eta} \int \Bigg\{ \NL
x_jp_i\NR(\varphi)\mathcal{H}(\varphi-\epsilon)\nonumber\\
&\qquad\qquad 
-\bigg( x_i(\varphi+\epsilon) +\tilde{\epsilon} x_{i}'
(\varphi+\epsilon) + \frac{1}{2}\tilde{\epsilon}^{2}x_{i}''
(\varphi+\epsilon)\bigg)\bigg( \NL p_{j}\mathcal{H} \NR  (\varphi)
-\frac{1}{\eta\tilde{\epsilon}^{2}}p_{j}(\varphi)\bigg)\nonumber\\
&\qquad\qquad  +
x_i(\varphi+\epsilon)\bigg(
\NL x'_j \zeta' \NR (\varphi) -\frac{1}{\eta\tilde{\epsilon}^{2}}
p_{j} (\varphi) \bigg) 
-x_i'(\varphi+\epsilon)
\bigg( \NL x_{j}\zeta '\NR (\varphi) 
+\frac{1}{\eta\tilde{\epsilon}} p_{j} (\varphi)
\bigg)\Bigg\} d \varphi + \mathcal{O} (\tilde{\epsilon})\\
& = 
\frac{2\pi i}{\eta} \int \Bigg\{ \NL
x_jp_i\NR(\varphi+\epsilon)\mathcal{H}(\varphi)
-x_i(\varphi+\epsilon) \NL p_{j}\mathcal{H} \NR(\varphi)
\nonumber\\
&\qquad\qquad
+ x_i(\varphi+\epsilon)\NL x'_j \zeta' \NR (\varphi)
-x_i'(\varphi+\epsilon) \NL x_{j}\zeta '\NR (\varphi)
+\frac{1}{2\eta }x_{i}''(\varphi+\epsilon)p_{j}(\varphi)
\Bigg\} d \varphi + \mathcal{O} (\tilde{\epsilon}) \ ,
\end{align}
where in the first summand we shifted the integration variable by
$\epsilon$. As expected there is no singularity when $\tilde{\epsilon}$ is taken
to zero. 

We now want to analyse the possible singularities in $\epsilon$. The
singularities between normal-ordered expressions arise from
singularities between the constituents, we have e.g.\ 
\begin{align}
x_{i} (\varphi+\epsilon) \NL x'_j \zeta' \NR (\varphi)
&= \frac{2\pi}{\eta}x_{i} (\varphi+\epsilon) \NL x'_j \NL p\cdot x' \NR  \NR (\varphi)\\
&= \frac{1}{\eta \epsilon} \NL x_{j}' p_{i}\NR (\varphi) +
\text{regular} \ .
\end{align}
OPE normal ordering is in general not associative, and we want to
define the normal-ordering of several operators in a right-nested way,
\begin{equation}
\NL ABC \NR := \NL A \NL BC\NR \NR \ .
\end{equation}
In the case at hand we have (see appendix~\ref{sec:appendix-non-assoc})
\begin{align}
\NL \NL x_{j}p_{i} \NR \mathcal{H} \NR &= \NL x_{j}p_{i}\mathcal{H}\NR
-\frac{1}{2\eta} \big( \NL x_{j}'' p_{i}\NR -2 \NL p_{i}'
x_{j}'\NR  \big)\\
\NL \NL x_{i}x_{j}'\NR \zeta' \NR &= \NL  x_{i}x_{j}' \zeta' \NR
+\frac{1}{2\eta} \big(2\NL x_{j}''p_{i}\NR - \NL x_{i}''p_{j}\NR
\big) \ ,
\end{align}
so that we find
\begin{align}
C_{ij}^{x\mathcal{H},x\mathcal{H}}(\epsilon,0) & =
\frac{2\pi i}{\eta} \int \Bigg\{ \bigg(
\NL \NL x_{j}p_{i}\NR \mathcal{H}\NR (\varphi) 
-\frac{1}{\epsilon^{2}\eta} \NL p_{i}x_{j}\NR
(\varphi) \bigg)  
-\bigg( \NL x_{i}p_{j}\mathcal{H} \NR (\varphi)
+\frac{1}{\epsilon\eta} \NL p_{j} x_{i}' \NR (\varphi)\bigg)
\nonumber\\
&\qquad\quad
+ \bigg( \NL x_{i} x_{j}' \zeta' \NR (\varphi)
+\frac{1}{\epsilon\eta} \NL x_{j}' p_{i}\NR (\varphi)\bigg)
- \bigg( \NL x_{i}' x_{j} \zeta' \NR (\varphi)
-\frac{1}{\epsilon^{2}\eta} \NL x_{j} p_{i} \NR (\varphi)\bigg) 
+\frac{1}{2\eta}\NL x_{i}'' p_{j}\NR (\varphi)
\Bigg\} d \varphi + \mathcal{O} (\epsilon)\nonumber\\
&= \frac{2\pi i}{\eta} \int \Bigg\{
\frac{1}{\epsilon\eta}\bigg(\NL p_{i}x_{j}'\NR (\varphi) -\NL
p_{j}x_{i}'\NR (\varphi)\bigg) + \frac{1}{2\eta}\NL x_{i}'' p_{j}\NR
(\varphi) -\frac{3}{2\eta} \NL x_{j}''p_{i}\NR (\varphi)
 \nonumber\\
&\qquad \qquad 
+\NL x_{j}p_{i}\mathcal{H}\NR (\varphi) 
-\NL x_{i}p_{j}\mathcal{H} \NR (\varphi)
+  \NL x_{i} x_{j}' \zeta' \NR (\varphi)
- \NL x_{i}' x_{j} \zeta' \NR (\varphi)
\Bigg\} d \varphi + \mathcal{O} (\epsilon)\ .
\label{CxHxH}
\end{align}
The remaining singularity in $\epsilon$ will be cancelled by the
contribution $C_{ij}^{x\mathcal{H},p\zeta}$ that we discuss in the following.

\subsection{Commutators of the form $[x_{i}\mathcal{H},p_{j}\zeta]$}

We now turn to the analysis of the contribution
\begin{equation}
C_{ij}^{x\mathcal{H},p\zeta}(\epsilon,\tilde{\delta}) = -\frac{1}{2}\int \int
\Big[
x_i(\varphi+\epsilon) \mathcal{H}(\varphi) ,  \zeta
(\tilde{\varphi}+\tilde{\delta}) p_j(\tilde{\varphi}) +p_{j}
(\tilde{\varphi}) \zeta (\tilde{\varphi}+\tilde{\delta})  \Big] d\varphi\,d\tilde{\varphi}\ .
\end{equation}
Evaluating the commutator we find (for $i\not= j$)
\begin{align}
C_{ij}^{x\mathcal{H},p\zeta}(\epsilon,\tilde{\delta}) &=
\frac{2\pi i}{\eta}\int \Big(x_i(\varphi+\epsilon)
p_j(\varphi-\tilde{\delta})\mathcal{H}(\varphi)-
x_i(\varphi+\epsilon-\tilde{\delta}) x_j'(\varphi-\tilde{\delta})\zeta'(\varphi) \Big) d\varphi\nonumber \\
&\quad +\frac{i}{\eta}\int\! \int x_i'(\varphi+\epsilon) p_j(\tilde{\varphi})\mathcal{H}(\varphi)\, \partial_{\varphi}\big(G(\tilde{\varphi}+\tilde{\delta},\varphi+\epsilon)-G(\tilde{\varphi}+\tilde{\delta},\varphi)\big)d\varphi \,d \tilde{\varphi} \nonumber\\
&\quad -\frac{\pi i}{3\eta^2} (D-2) \int x''_i(\varphi+\tilde{\delta} +
\epsilon ) p_j(\varphi) \,d\varphi \ .
\end{align}
We expand this expression first in $\tilde{\delta}$, and we obtain
\begin{align}
C_{ij}^{x\mathcal{H},p\zeta}(\epsilon,\tilde{\delta}) &=
\frac{2\pi i}{\eta}\int \Bigg(x_i(\varphi+\epsilon)
 \bigg( \NL p_{j}\mathcal{H}\NR (\varphi)
 -\frac{1}{\tilde{\delta}^{2}\eta} p_{j}(\varphi)\bigg) \nonumber\\
&\qquad \qquad 
-\bigg( x_{i} (\varphi+\epsilon)
-\tilde{\delta}x_{i}'(\varphi+\epsilon)+\frac{\tilde{\delta}^{2}}{2}x_{i}''(\varphi+\epsilon)\bigg)
\bigg( \NL x_{j}'\zeta'\NR (\varphi) -\frac{1}{\tilde{\delta}^{2}\eta} p_{j} (\varphi) \bigg)\Bigg) d\varphi\nonumber \\
&\quad +\frac{i}{\eta}\int\! \int x_i'(\varphi+\epsilon) \bigg( p_j(\tilde{\varphi})\mathcal{H}(\varphi)-\frac{2\pi}{\eta}S_{\text{sing}} (\tilde{\varphi},\varphi)p_{j} (\varphi)\bigg)\, \partial_{\varphi}\big(G(\tilde{\varphi}+\tilde{\delta},\varphi+\epsilon)-G(\tilde{\varphi}+\tilde{\delta},\varphi)\big)d\varphi \,d \tilde{\varphi} \nonumber\\
&\quad +\frac{i}{\eta}\int\! \int x_i'(\varphi+\epsilon) \bigg( \frac{2\pi}{\eta}S_{\text{sing}} (\tilde{\varphi},\varphi)p_{j}(\varphi)\bigg)\, \partial_{\varphi}\big(G(\tilde{\varphi}+\tilde{\delta},\varphi+\epsilon)-G(\tilde{\varphi}+\tilde{\delta},\varphi)\big)d\varphi \,d \tilde{\varphi} \nonumber\\
&\quad -\frac{\pi i}{3\eta^2} (D-2) \int x''_i(\varphi+
\epsilon ) p_j(\varphi) \,d\varphi + \mathcal{O} (\tilde{\delta})\ .
\label{CxHpzexpansion}
\end{align}
In the third and fourth line we have subtracted and added the singular
piece of $p_{j}(\tilde{\varphi})\mathcal{H}(\varphi)$. In the third line, there
is therefore no singularity coming from the operator part when
$\varphi$ and $\tilde{\varphi}$ are close together, and we can just
set $\tilde{\delta}$ to $0$. The fourth line can be evaluated by
writing $S_{\text{sing}}$ as a derivative and then using partial integration,
\begin{align}
&\frac{i}{\eta}\int\! \int x_i'(\varphi+\epsilon) \bigg(
\frac{2\pi}{\eta}S_{\text{sing}}
(\tilde{\varphi},\varphi)p_{j}(\varphi)\bigg)\,
\partial_{\varphi}\big(G(\tilde{\varphi}+\tilde{\delta},\varphi+\epsilon)-G(\tilde{\varphi}+\tilde{\delta},\varphi)\big)d\varphi
\,d \tilde{\varphi}\nonumber\\
&\quad = \frac{i}{2\eta^{2}}\int\! \int x_i'(\varphi+\epsilon)
\partial_{\tilde{\varphi}} \bigg( \cos
\frac{\tilde{\varphi}-\varphi}{2} \mathcal{P}\frac{1}{\sin
\frac{\tilde{\varphi}-\varphi}{2}}\bigg)
p_{j}(\varphi)\,
\partial_{\varphi}\big(G(\tilde{\varphi}+\tilde{\delta},\varphi+\epsilon)-G(\tilde{\varphi}+\tilde{\delta},\varphi)\big)d\varphi
\,d \tilde{\varphi}\\
&\quad = -\frac{i}{2\eta^{2}}\int\! \int x_i'(\varphi+\epsilon)
\bigg( \cos
\frac{\tilde{\varphi}-\varphi}{2} \mathcal{P}\frac{1}{\sin
\frac{\tilde{\varphi}-\varphi}{2}}\bigg)
p_{j}(\varphi)\,
\big( -2\pi \delta(\tilde{\varphi}-\varphi+\tilde{\delta}-\epsilon)
+2\pi \delta (\tilde{\varphi}-\varphi+\tilde{\delta})
\big)d\varphi
\,d \tilde{\varphi}\\
&\quad = \frac{i\pi}{\eta^{2}}\int x_i'(\varphi+\epsilon)p_{j}(\varphi)
\bigg( \cos
\frac{\epsilon-\tilde{\delta}}{2} \frac{1}{\sin
\frac{\epsilon-\tilde{\delta}}{2}}
+\cos
\frac{\tilde{\delta}}{2} \frac{1}{\sin
\frac{\tilde{\delta}}{2}}
\bigg)
d\varphi\\
&\quad = \frac{i\pi}{\eta^{2}}\int x_i'(\varphi+\epsilon)p_{j}(\varphi)
\bigg( \cos
\frac{\epsilon}{2} \frac{1}{\sin
\frac{\epsilon}{2}}
+\frac{2}{\tilde{\delta}}
\bigg)
d\varphi + \mathcal{O} (\tilde{\delta}) \ .
\end{align}
Inserting this result into~\eqref{CxHpzexpansion} we obtain
\begin{align}
C_{ij}^{x\mathcal{H},p\zeta}(\epsilon,\tilde{\delta}) &=
\frac{2\pi i}{\eta}\int \Bigg(x_i(\varphi+\epsilon)
 \NL p_{j}\mathcal{H}\NR (\varphi) -  x_{i} (\varphi+\epsilon) \NL
 x_{j}'\zeta'\NR (\varphi)
+ \frac{1}{2\eta}x_{i}''(\varphi+\epsilon)p_{j}(\varphi)
\Bigg) d\varphi\nonumber \\
&\quad +\frac{i}{\eta}\int\! \int x_i'(\varphi+\epsilon) \bigg( p_j(\tilde{\varphi})\mathcal{H}(\varphi)-\frac{2\pi}{\eta}S_{\text{sing}} (\tilde{\varphi},\varphi)p_{j} (\varphi)\bigg)\, \partial_{\varphi}\big(G(\tilde{\varphi},\varphi+\epsilon)-G(\tilde{\varphi},\varphi)\big)d\varphi \,d \tilde{\varphi} \nonumber\\
&\quad  +\frac{i\pi}{\eta^{2}}\int x_i'(\varphi+\epsilon)p_{j}(\varphi)
\bigg( \cos
\frac{\epsilon}{2} \frac{1}{\sin
\frac{\epsilon}{2}}
\bigg)
d\varphi\nonumber\\
&\quad -\frac{\pi i}{3\eta^2} (D-2) \int x''_i(\varphi+
\epsilon ) p_j(\varphi) \,d\varphi + \mathcal{O} (\tilde{\delta})\ .
\end{align}
We observe that the expression contains no singularity in
$\tilde{\delta}$. Now we expand in $\epsilon$,
\begin{align}
C_{ij}^{x\mathcal{H},p\zeta}(\epsilon,0) &=
\frac{2\pi i}{\eta}\int \Bigg(
\NL x_{i}p_{j}\mathcal{H}\NR (\varphi) + \frac{1}{\epsilon\eta} \NL
p_{j}x_{i}' \NR (\varphi)
- \NL x_{i}x_{j}'\zeta'\NR (\varphi) -\frac{1}{\epsilon\eta}\NL
p_{i}x_{j}'\NR (\varphi)
+ \frac{1}{2\eta}\NL x_{i}''p_{j}\NR (\varphi)
\Bigg) d\varphi\nonumber \\
&\quad +\frac{i}{\eta}
\int \bigg(-\frac{2\pi}{\epsilon\eta}\NL x_{i}'p_{j}\NR(\varphi)
+\frac{\pi}{\eta}\NL x_{i}''p_{j}\NR (\varphi) \bigg)d\varphi\nonumber\\
&\quad  +\frac{i\pi}{\eta^{2}}\int \bigg(\frac{2}{\epsilon}
\NL x_{i}'p_{j}\NR (\varphi) + 2 \NL x_{i}''p_{j}\NR (\varphi)
\bigg)
d\varphi\nonumber\\
&\quad -\frac{\pi i}{3\eta^2} (D-2) \int \NL x''_i p_j\NR (\varphi)
\,d\varphi + \mathcal{O} (\epsilon)\\
&=\frac{2\pi i}{\eta} \int \Bigg\{ \frac{1}{\epsilon\eta}\bigg( 
\NL p_{j}x_{i}' \NR (\varphi)-\NL p_{i}x_{j}'\NR (\varphi)
\bigg) +
\NL x_{i}p_{j}\mathcal{H}\NR (\varphi) 
- \NL x_{i}x_{j}'\zeta'\NR (\varphi) 
\nonumber \\
&\qquad \qquad \quad 
+\frac{2}{\eta}
\NL x_{i}''p_{j}\NR (\varphi) 
-\frac{D-2}{6\eta} \NL x''_i p_j\NR (\varphi) \Bigg\}
d\varphi + \mathcal{O} (\epsilon)\ .
\end{align}
The remaining singularity cancels the terms that we found in
$C_{ij}^{x\mathcal{H},x\mathcal{H}}$ (see~\eqref{CxHxH}), so that we
find
\begin{align}
&\lim_{\epsilon \to 0} \int 
\Big[ x_i(\varphi+\epsilon) \mathcal{H}(\varphi) ,
M_{j-}\Big]d\varphi \nonumber\\
&\qquad =\lim_{\epsilon\to 0}
\Big(C_{ij}^{x\mathcal{H},x\mathcal{H}}(\epsilon,0) + C_{ij}^{x\mathcal{H},p\zeta}(\epsilon,0) \Big)\\
&\qquad = \frac{2\pi i}{\eta} \int \Bigg\{
\NL x_{j}p_{i}\mathcal{H}\NR (\varphi) 
- \NL x_{i}' x_{j} \zeta' \NR (\varphi)
+\frac{5}{2\eta}
\NL x_{i}''p_{j}\NR (\varphi)
-\frac{3}{2\eta} \NL x_{j}''p_{i}\NR (\varphi)
-\frac{D-2}{6\eta} \NL x''_i p_j\NR (\varphi)
\Bigg\}
d\varphi \ .
\end{align}

\subsection{Remaining commutators and final result}

The remaining commutators can be computed analogously to the
computations we displayed above, which is done in
the appendices~\ref{sec:commutatorpzetaxH}
and~\ref{sec:commutatorpzetapzeta}, and the results are given
in~\eqref{resultC_pzeta_xH} and~\eqref{resultC_pzeta_pzeta}. Their sum does not contain any
singularity, and one obtains
\begin{align}
&\lim_{\delta \to 0} \int \bigg[-\frac{1}{2}
\big(\zeta(\varphi+\delta)p_{i}(\varphi)+p_{i}(\varphi)\zeta(\varphi+\delta)\big),
 M_{j-}\bigg] d\varphi\nonumber\\
&\qquad = \lim_{\delta \to 0} \big(
C_{ij}^{p\zeta,x\mathcal{H}}(\delta,0)+C_{ij}^{p\zeta,p\zeta}(\delta,0)\big)\\
&\qquad =\frac{2\pi i}{\eta}\int
\Bigg\{ 
-\NL p_{i}x_{j}\mathcal{H}\NR (\varphi)
+\NL x_{j}x_{i}'\zeta '\NR (\varphi)
 -\frac{5}{2\eta}\NL x_{j}'' p_{i}\NR (\varphi)+\frac{3}{2\eta} \NL p_{j}x_{i}''\NR (\varphi)
+\frac{D-2}{6\eta}\NL x_{j}'' p_{i}\NR (\varphi) 
\Bigg\} d\varphi \ .
\end{align}
Combining now all contributions we find for the quantum commutator the
final result
\begin{equation}
\Big[M_{i-},M_{j-} \Big] = \frac{\pi i}{3\eta^{2}} (D-26) \int \Big(
\NL x_j''p_{i}\NR (\varphi) - \NL x_{i}''p_{j}\NR (\varphi) \Big)
d\varphi \ ,
\end{equation}
which vanishes if $D=26$.

The dimension-dependent
term (linear in $D-2$) came from the commutator of $\mathcal{H}$ and
$\zeta$, the other anomalous terms came from all commutators that
occur in the computation. As already remarked in the introduction,
we did not introduce a normal-ordering constant to define
$\mathcal{H}$. In principle such a shift ($\mathcal{H}\to\mathcal{H}+\frac{\text{const.}}{\eta}$)
could be considered (then
one would derive from demanding a vanishing commutator
$[M_{i-},M_{j-}]$ that this constant is zero),
but the definition of $\mathcal{H}$ via
OPE normal ordering on the cylinder appears to be most natural.

\subsection*{Acknowledgements}
We would like to thank J{\"u}rg Fr{\"o}hlich and Volker Schomerus, AEI and KTH,
and the Swedish Research Council.

\appendix

\section{Commutators of the form $[p_{i}\zeta,x_{j}\mathcal{H}]$}
\label{sec:commutatorpzetaxH}
We discuss here the contribution 
\begin{equation}
C_{ij}^{p\zeta,x\mathcal{H}}(\delta,\tilde{\epsilon}) = -\frac{1}{2}\int \int  
 \Big[\zeta
(\varphi+\delta) p_i(\varphi) +p_{i}
(\varphi) \zeta (\varphi+\delta),
 x_j(\tilde{\varphi}+\tilde{\epsilon}) \mathcal{H}(\tilde{\varphi})
 \Big]d\varphi\,d\tilde{\varphi}\ .
\end{equation}
For $i\not =j$ the commutator is given by
\begin{align}
C_{ij}^{p\zeta,x\mathcal{H}}(\delta,\tilde{\epsilon})
&=
-\frac{2\pi i}{\eta}\int \Big(x_j(\varphi+\tilde{\epsilon})
p_i(\varphi-\delta)\mathcal{H}(\varphi)-
x_j(\varphi+\tilde{\epsilon}-\delta) x_i'(\varphi-\delta)\zeta'(\varphi) \Big) d\varphi\nonumber \\
&\quad -\frac{i}{\eta}\int\! \int x_j'(\tilde{\varphi}+\tilde{\epsilon}) p_i(\varphi)\mathcal{H}(\tilde{\varphi})\, \partial_{\tilde{\varphi}}\big(G(\varphi+\delta,\tilde{\varphi}+\tilde{\epsilon})-G(\varphi+\delta,\tilde{\varphi})\big)d\varphi \,d \tilde{\varphi} \nonumber\\
&\quad +\frac{\pi i}{3\eta^2} (D-2) \int x''_j(\varphi+\delta +
\tilde{\epsilon} ) p_i(\varphi) \,d\varphi \ .
\end{align}
The result is regular when we expand in $\tilde{\epsilon}$, and we
obtain
\begin{align}
C_{ij}^{p\zeta,x\mathcal{H}}(\delta,0) &=
\frac{2\pi i}{\eta}\int
\Bigg\{ -p_{i} (\varphi-\delta) \NL x_{j}\mathcal{H}\NR (\varphi) 
+ \NL x_{j}x_{i}'\NR (\varphi-\delta) \zeta' (\varphi)\nonumber\\
&\qquad \qquad -\frac{1}{2\eta}x_{j}''(\varphi+\delta)p_{i}(\varphi)
+\frac{D-2}{6\eta}x_{j}''(\varphi+\delta)p_{i}(\varphi) 
\Bigg\} d\varphi\ .
\end{align}
Expanding in $\delta$ we find
\begin{align}
C_{ij}^{p\zeta,x\mathcal{H}}(\delta,0) &=
\frac{2\pi i}{\eta}\int
\Bigg\{ \frac{1}{\delta\eta} \bigg(\NL p_{i}x_{j}'\NR (\varphi) 
- \NL p_{j}x_{i}'\NR (\varphi)  \bigg)
-\NL p_{i}x_{j}\mathcal{H}\NR (\varphi)
+\NL \NL x_{j}x_{i}'\NR \zeta '\NR (\varphi)
\nonumber\\
&\qquad \qquad -\frac{1}{2\eta}\NL x_{j}'' p_{i}\NR (\varphi)
+\frac{D-2}{6\eta}\NL x_{j}'' p_{i}\NR (\varphi) 
\Bigg\} d\varphi + \mathcal{O} (\delta)\\
 &=
\frac{2\pi i}{\eta}\int
\Bigg\{ \frac{1}{\delta\eta} \bigg(\NL p_{i}x_{j}'\NR (\varphi) 
- \NL p_{j}x_{i}'\NR (\varphi)  \bigg)
-\NL p_{i}x_{j}\mathcal{H}\NR (\varphi)
+\NL x_{j}x_{i}' \zeta '\NR (\varphi)
\nonumber\\
&\qquad \qquad +\frac{1}{\eta}\NL x_{i}''p_{j}\NR -\frac{1}{\eta}\NL x_{j}'' p_{i}\NR (\varphi)
+\frac{D-2}{6\eta}\NL x_{j}'' p_{i}\NR (\varphi) 
\Bigg\} d\varphi + \mathcal{O} (\delta)
\ .
\label{resultC_pzeta_xH}
\end{align}

\section{Commutators of the form $[p_{i}\zeta,p_{j}\zeta]$}
\label{sec:commutatorpzetapzeta}

Let us now discuss the term
\begin{equation}
C_{ij}^{p\zeta,p\zeta}(\delta,\tilde{\delta})
= \frac{1}{4}  \int \int  
 \Big[\zeta
(\varphi+\delta) p_i(\varphi) +p_{i}
(\varphi) \zeta (\varphi+\delta),
\zeta
(\tilde{\varphi}+\tilde{\delta}) p_j(\tilde{\varphi}) +p_{j}
(\tilde{\varphi}) \zeta (\tilde{\varphi}+\tilde{\delta})
\Big] d\varphi\,d\tilde{\varphi}\ .
\end{equation}
One can straightforwardly show that the four different terms that one
obtains from expanding the commutator above all lead to the same contribution,
\begin{align}
C_{ij}^{p\zeta,p\zeta}(\delta,\tilde{\delta}) &= \frac{1}{4}  \int \int 
 \Big[\big[ \zeta
(\varphi+\delta), p_i(\varphi)\big] + 2p_{i}
(\varphi) \zeta (\varphi+\delta),
\big[ \zeta
(\tilde{\varphi}+\tilde{\delta}), p_j(\tilde{\varphi})\big] + 2 p_{j}
(\tilde{\varphi}) \zeta (\tilde{\varphi}+\tilde{\delta})
\Big] d\varphi\,d\tilde{\varphi}\\
& =  \int \int
\Big[p_{i}
(\varphi) \zeta (\varphi+\delta),
p_{j} (\tilde{\varphi}) \zeta (\tilde{\varphi}+\tilde{\delta})
\Big] d\varphi\,d\tilde{\varphi}\ .
\end{align}
By explicitly evaluating the commutator we find 
\begin{align}
C_{ij}^{p\zeta,p\zeta}(\delta,\tilde{\delta}) &=
\frac{2 \pi i}{\eta}\int \Big( p_{i} (\varphi-\delta-\tilde{\delta})
p_{j} (\varphi-\tilde{\delta}) - p_{j} (\varphi-\delta
-\tilde{\delta}) p_{i} (\varphi-\delta)\Big)\zeta (\varphi) \,d\varphi\nonumber\\
& \quad + \frac{i}{\eta}\int \int \Big\{p_{i} (\varphi-\delta)p_{j}'
(\tilde{\varphi}-\tilde{\delta})\partial_{\tilde{\varphi}}\big(G
(\varphi,\tilde{\varphi}-\tilde{\delta})-G (\varphi
,\tilde{\varphi}) \big)\zeta (\tilde{\varphi})\nonumber\\
&\qquad \qquad \quad -p_{j} (\tilde{\varphi}-\tilde{\delta}) p_{i}' (\varphi-\delta)
\partial_{\varphi}\big(G
(\tilde{\varphi},\varphi-\delta)-G
(\tilde{\varphi},\varphi) \big)\zeta (\varphi)
\Big\}d\varphi \,d\tilde{\varphi} \\
&=  \frac{i}{\eta}\int \int \Big\{-p_{i} (\varphi-\delta)p_{j}
(\tilde{\varphi}-\tilde{\delta})\partial_{\tilde{\varphi}}\big(G
(\varphi,\tilde{\varphi}-\tilde{\delta})-G (\varphi
,\tilde{\varphi}) \big)\zeta '(\tilde{\varphi})\nonumber\\
&\qquad \qquad \quad +p_{j} (\tilde{\varphi}-\tilde{\delta}) p_{i} (\varphi-\delta)
\partial_{\varphi}\big(G
(\tilde{\varphi},\varphi-\delta)-G
(\tilde{\varphi},\varphi) \big)\zeta' (\varphi)
\Big\}d\varphi \,d\tilde{\varphi} \ .
\end{align}
We now expand this expression in $\tilde{\delta}$ analogously to the cases we
discussed before. We obtain
\begin{align}
C_{ij}^{p\zeta,p\zeta}(\delta,\tilde{\delta}) &=
\frac{i}{\eta}\int \int \Bigg\{-p_{i} (\varphi-\delta)
\bigg( -\frac{1}{\eta\tilde{\delta}^{2}}x_{j}'(\tilde{\varphi}) 
+ \NL p_{j}\zeta'\NR (\tilde{\varphi})\bigg)
\bigg(-\tilde{\delta}\partial^{2}_{\tilde{\varphi}}G
(\varphi,\tilde{\varphi}) +
\frac{\tilde{\delta}^{2}}{2}\partial^{3}_{\tilde{\varphi}}G
(\varphi,\tilde{\varphi})
 \bigg)
\nonumber\\
&\qquad \qquad \quad 
+p_{i} (\varphi-\delta) 
\partial_{\varphi}\big(G
(\tilde{\varphi},\varphi-\delta)-G
(\tilde{\varphi},\varphi) \big)
\bigg( p_{j} (\tilde{\varphi}-\tilde{\delta}) \zeta'
(\varphi)-\frac{2\pi}{\eta}S_{\text{sing}}
(\tilde{\varphi}-\tilde{\delta},\varphi)x_{j}'(\varphi)\bigg)
\nonumber\\
&\qquad \qquad \quad 
+p_{i} (\varphi-\delta) 
\partial_{\varphi}\big(G
(\tilde{\varphi},\varphi-\delta)-G
(\tilde{\varphi},\varphi) \big)
\frac{2\pi}{\eta}S_{\text{sing}}
(\tilde{\varphi}-\tilde{\delta},\varphi)x_{j}'(\varphi)
\Bigg\}d\varphi \,d\tilde{\varphi} + \mathcal{O} (\tilde{\delta})\\
&=
\frac{\pi i}{\eta^{2}}\int \bigg(-\frac{2}{\tilde{\delta}}p_{i} (\varphi-\delta)
x_{j}'(\varphi) - p_{i} (\varphi-\delta)x_{j}''(\varphi)\bigg) d\varphi
\nonumber\\
&\quad 
+\frac{i}{\eta}\int \int p_{i} (\varphi-\delta) 
\partial_{\varphi}\big(G
(\tilde{\varphi},\varphi-\delta)-G
(\tilde{\varphi},\varphi) \big)
\bigg( p_{j} (\tilde{\varphi}) \zeta'
(\varphi)-\frac{2\pi}{\eta}S_{\text{sing}}
(\tilde{\varphi},\varphi)x_{j}'(\varphi)\bigg)d\varphi \,d\tilde{\varphi}
\nonumber\\
&\quad 
+\frac{\pi i}{\eta^{2}}\int p_{i}(\varphi-\delta) x_{j}'(\varphi) \bigg(
-\cos \frac{\delta+\tilde{\delta}}{2} \frac{1}{\sin
\frac{\delta+\tilde{\delta}}{2}} +
\cos\frac{\tilde{\delta}}{2}\frac{1}{\sin\frac{\tilde{\delta}}{2}} \bigg)
d\varphi + \mathcal{O} (\tilde{\delta})\\
&=
\frac{\pi i}{\eta^{2}}\int \bigg( - p_{i} (\varphi-\delta)x_{j}''(\varphi)\bigg)d\varphi
\nonumber\\
&\quad 
+\frac{i}{\eta}\int \int p_{i} (\varphi-\delta) 
\partial_{\varphi}\big(G
(\tilde{\varphi},\varphi-\delta)-G
(\tilde{\varphi},\varphi) \big)
\bigg( p_{j} (\tilde{\varphi}) \zeta'
(\varphi)-\frac{2\pi}{\eta}S_{\text{sing}}
(\tilde{\varphi},\varphi)x_{j}'(\varphi)\bigg)d\varphi \,d\tilde{\varphi}
\nonumber\\
&\quad 
+\frac{\pi i}{\eta^{2}}\int p_{i}(\varphi-\delta) x_{j}'(\varphi) \bigg(
-\cos \frac{\delta+\tilde{\delta}}{2} \frac{1}{\sin
\frac{\delta+\tilde{\delta}}{2}} \bigg)
d\varphi + \mathcal{O} (\tilde{\delta}) \ .
\end{align}
As expected there is no singularity in $\tilde{\delta}$. We now set
$\tilde{\delta}=0$ and expand in $\delta$,
\begin{align}
C_{ij}^{p\zeta,p\zeta}(\delta,0) &=
\frac{\pi i}{\eta^{2}}\int \bigg( -\NL p_{i} x_{j}''\NR (\varphi)\bigg) d\varphi
\nonumber\\
&\quad 
+\frac{\pi i}{\eta^{2}}\int \bigg(
\frac{2}{\delta}\NL p_{j}x_{i}'\NR (\varphi) + \NL p_{j}x_{i}''\NR
(\varphi) \bigg)d\varphi
\nonumber\\
&\quad 
+\frac{\pi i}{\eta^{2}}\int  \bigg(
-\frac{2}{\delta}\NL p_{i}x_{j}'\NR (\varphi)
-2 \NL p_{i}x_{j}''\NR (\varphi) \bigg)
d\varphi + \mathcal{O}(\delta)\\
&=
\frac{2\pi i}{\eta}\int \Bigg\{ 
\frac{1}{\delta\eta} \bigg(
\NL p_{j}x_{i}'\NR (\varphi) - \NL p_{i}x_{j}'\NR (\varphi)
\bigg)\nonumber\\
&\qquad \qquad \quad + \bigg(
-\frac{3}{2\eta}\NL p_{i}x_{j}''\NR (\varphi) 
+ \frac{1}{2\eta} \NL p_{j}x_{i}''\NR (\varphi)\bigg) 
\Bigg\} d\varphi + \mathcal{O}(\delta) \ .
\label{resultC_pzeta_pzeta}
\end{align}

\section{Non-associativity}
\label{sec:appendix-non-assoc}
OPE normal ordering is in general not associative. In this appendix we
will discuss those cases that are relevant in the main text.

The first identity, we want to explain is
\begin{equation}\label{app:firstnonassoc}
\NL \NL x_{j}p_{i} \NR \mathcal{H} \NR = \NL x_{j}p_{i}\mathcal{H}\NR
-\frac{1}{2\eta} \big( \NL x_{j}'' p_{i}\NR -2 \NL p_{i}'
x_{j}'\NR  \big) \ .
\end{equation}
The simplest way to show this is to write 
\begin{equation}
\mathcal{H} = \sum_{k} \mathcal{H}_{k} \quad ,\quad 
\mathcal{H}_{k} = \frac{\pi}{\eta} \Big(\NL p_{k}^{2}\NR +\NL
x_{k}'^{2}\NR  \Big)\ ,
\end{equation}
and consider the summands $\mathcal{H}_{k}$ individually. For $k$
different from $i$ and $j$ there is no singularity, and thus normal
ordering of $x_{j}$, $p_{i}$ and $\mathcal{H}_{k}$ is associative. Now
consider $k=i$,
\begin{align}
\NL \NL x_{j}p_{i} \NR \mathcal{H}_{i} \NR (\varphi) &= 
\lim_{\epsilon \to 0} \bigg( x_{j} (\varphi-\epsilon)
p_{i}(\varphi-\epsilon) \mathcal{H}_{i} (\varphi)  
+\frac{1}{\epsilon^{2}\eta} \Big( x_{j}(\varphi)-\epsilon
x_{j}'(\varphi)\Big)p_{i} (\varphi)
\bigg)\\
&= \lim_{\epsilon\to 0} \bigg( x_{j} (\varphi-\epsilon)
\Big(\NL p_{i}\mathcal{H}_{i}\NR (\varphi)  -\frac{1}{\epsilon^{2}\eta}p_{i}(\varphi)\Big)  
+\frac{1}{\epsilon^{2}\eta} \Big( x_{j}(\varphi)-\epsilon x_{j}'(\varphi)\Big)p_{i} (\varphi)
\bigg)\\
&=\NL x_{j}p_{i}\mathcal{H}_{i}\NR (\varphi) -\frac{1}{2\eta} \NL x_{j}''p_{i}\NR (\varphi) \ .
\label{app:xjpiHi}
\end{align}
If we instead consider $k=j$, we find
\begin{align}
\NL \NL x_{j}p_{i} \NR \mathcal{H}_{j} \NR (\varphi) &= 
\lim_{\epsilon \to 0} \bigg( x_{j} (\varphi-\epsilon)
p_{i}(\varphi-\epsilon) \mathcal{H}_{j} (\varphi)  
+ \frac{1}{\epsilon\eta}x_{j}'(\varphi)p_{i}(\varphi)
\bigg)\\
&= \lim_{\epsilon\to 0} \bigg(x_{j} (\varphi-\epsilon) \Big( p_{i}
(\varphi)-\epsilon p_{i}' (\varphi)\Big) \mathcal{H}_{j}(\varphi) 
 + \frac{1}{\epsilon\eta}x_{j}'(\varphi)p_{i}(\varphi)
 \bigg)\\
&=\lim_{\epsilon\to 0} \bigg(\Big(\NL x_{j}p_{i}\mathcal{H}_{j}\NR
(\varphi) -\frac{1}{\epsilon\eta}x_{j}'(\varphi)p_{i}(\varphi) \Big) 
+\frac{1}{\eta}x_{j}'(\varphi)p_{i}'(\varphi)
+\frac{1}{\epsilon\eta}x_{j}'(\varphi)p_{i}(\varphi)
\bigg)\\
&=\NL x_{j}p_{i}\mathcal{H}_{j}\NR (\varphi) +\frac{1}{\eta}\NL
x_{j}'p_{i}'\NR (\varphi) \ .
\label{app:xjpiHj}
\end{align}
Combining~\eqref{app:xjpiHi} and~\eqref{app:xjpiHj} we arrive at the
desired result~\eqref{app:firstnonassoc}.

The second relation that we need is
\begin{equation}\label{app:secondnonass}
\NL \NL x_{i}x_{j}'\NR \zeta' \NR = \NL  x_{i}x_{j}' \zeta' \NR
+\frac{1}{2\eta} \big(2\NL x_{j}''p_{i}\NR - \NL x_{i}''p_{j}\NR
\big) \ .
\end{equation}
We can prove it analogously. Write 
\begin{equation}
\zeta' = \sum_{k} \zeta_{k}' \quad ,\quad \zeta_{k}'= \frac{2\pi}{\eta}\NL p_{k}x_{k}\NR \ ,
\end{equation}
and consider first the case when $\zeta_{i}'$ appears in the
normal-ordered product,
\begin{align}
\NL \NL x_{i}x_{j}'\NR \zeta_{i}' \NR &= 
\lim_{\epsilon\to 0} \bigg(
x_{i}(\varphi-\epsilon)x_{j}'(\varphi-\epsilon) \zeta_{i}'(\varphi)
+\frac{1}{\epsilon\eta}x_{j}' (\varphi)p_{i}(\varphi)
\bigg)\\
&=\lim_{\epsilon\to 0} \bigg(
x_{i}(\varphi-\epsilon)\Big( x_{j}'(\varphi)-\epsilon x_{j}'' (\varphi)\Big) \zeta_{i}'(\varphi)
+\frac{1}{\epsilon\eta}x_{j}' (\varphi)p_{i}(\varphi)
\bigg)\\
&=\lim_{\epsilon\to 0} \bigg( \Big( \NL x_{i}x_{j}'\zeta_{i}'\NR
(\varphi)-\frac{1}{\epsilon\eta}x_{j}' (\varphi)p_{i} (\varphi)\Big) +
\frac{1}{\eta}x_{j}''(\varphi)p_{i}(\varphi)
+\frac{1}{\epsilon\eta}x_{j}' (\varphi)p_{i}(\varphi)
\bigg)\\
&=\NL x_{i}x_{j}'\zeta_{i}'\NR
(\varphi) + \frac{1}{\eta}\NL x_{j}''p_{i}\NR (\varphi) \ .
\label{app:xixjzi}
\end{align}
Now consider the case, when $\zeta_{j}'$ occurs,
\begin{align}
\NL \NL x_{i}x_{j}'\NR \zeta_{j}' \NR &= 
\lim_{\epsilon\to 0} \bigg(
x_{i}(\varphi-\epsilon)x_{j}'(\varphi-\epsilon) \zeta_{j}'(\varphi)
+\frac{1}{\epsilon^{2}\eta}\Big( x_{i} (\varphi)-\epsilon x_{i}'(\varphi)\Big)p_{j}(\varphi)
\bigg)\\
&=\lim_{\epsilon\to 0} \bigg(
x_{i}(\varphi-\epsilon) \Big( \NL x_{j}'\zeta_{j}'\NR (\varphi)-\frac{1}{\epsilon^{2}\eta}p_{j} (\varphi) \Big)
+\frac{1}{\epsilon^{2}\eta}\Big( x_{i} (\varphi)-\epsilon x_{i}'(\varphi)\Big)p_{j}(\varphi)
\bigg)\\
&=\NL x_{i}x_{j}'\zeta_{j}'\NR (\varphi) -\frac{1}{2\eta} \NL
x_{i}''p_{j}\NR (\varphi)\ .
\label{app:xixjzj}
\end{align}
Combining~\eqref{app:xixjzi} and~\eqref{app:xixjzj} we arrive at the
final result~\eqref{app:secondnonass}.

\end{document}